# Drivers and barriers to integrating shared micromobility with public transport

A latent class clustering analysis of adoption attitudes in the Netherlands

Nejc Geržinič[1]*, Mark van Hagen[2], Hussein Al-Tamimi[3], Niels van Oort[4], Dorine Duives[4]

[1] Postdoctoral researcher, Department of Transport & Planning, TU Delft, Delft, Netherlands
[2] Principal consultant, Dutch Railways (NS), Utrecht, Netherlands
[3] Propositions & Partnerships, Dutch Railways (NS), Utrecht, Netherlands
[4] Associate professor, Department of Transport & Planning, TU Delft, Delft, Netherlands
* corresponding author

## Abstract

Shared micromobility (SMM), including bicycles, e-bikes, scooters, etc., is often cited as a solution to the first and especially the last mile problem of public transport (PT), yet when implemented, they often do not get adopted by a broader travelling public. As behavioural adaption is largely related to peoples' attitudes and perceptions, we develop a behavioural framework based on the UTAUT2 framework to gain better understanding why individuals may (not) be willing to use SMM. Through an exploratory factor analysis (EFA) and a latent class cluster analysis (LCCA), we study the adoption potential of SMM and assess drivers and barriers as perceived by different user groups. Our findings uncover six user groups; *Shared mobility positives*, *Car-oriented sharing neutrals, Older apprehensive sharers, Young eager adopters, (Shared) Mobility avoiders* and *Skilled sharing sceptics*. The *Young eager adopters* and *Shared mobility positives* tend to be the most open to adopting SMM and able to do so. *Older apprehensive sharers* would like to, but find it difficult or dangerous to use, while *Skilled sharing sceptics* are capable and confident, but have limited intention of using it. *Car-oriented sharing neutrals* and *(Shared) Mobility avoiders* are most negative about SMM, finding it difficult to use and dangerous. Factors relating to technological savviness, ease-of-use, physical safety and societal perception seem to be the strongest adoption predictors. Younger, high-educated males are the group most likely and open to using SMM, while older individuals with lower incomes and a lower level of education tend to be the least likely.

## Keywords

Attitudinal statements, Latent class cluster analysis, Micromobility, Multimodal transport, Shared mobility

## 1 Introduction

With the continuous growth of the mobility demand, a sustainable transition within the transport sector remains a challenge. For distances beyond the reach of active modes, i.e. approximately 5km (Jonkeren & Huang, 2024), rail-based public transport (PT) is the most sustainable alternative to the private car (Brand et al., 2021) due to its lower emissions as well as greater energy and space efficiency. However, several challenges remain in making rail-based PT more attractive, one of them being the first/last mile problem: accessing the starting point of a PT trip and egressing the end can often be cumbersome and time-consuming, taking up as much as 50% of the total trip time (Krygsman et al., 2004), despite making up only a small fraction of the trip distance.

Many different modes are used to access/egress PT stops, yet the most common are active modes, namely walking and (in certain countries also) cycling (Keijer & Rietveld, 2000; Ton et al., 2020). Cycling in particular is well suited as an access/egress mode to rail-based PT, increasing the range of walking, while being vastly more flexible than local PT, i.e. buses (Kager et al., 2016). In countries like the Netherlands, where cycling is commonplace, promoting and supporting it as a means of reaching the train station on the home-end of the trip has been highly successful. Large bicycle parking garages, good integration with train stations and existing high-quality cycling infrastructure have resulted in 39% of all train travellers using their bicycle to travel from their home to the station (with up to 60% in some cities), followed by walking (26%) and local PT (24%). On the activity-end of the trip however, cycling only holds a 13% share, behind both walking (52%) and local PT (28%) (Schakenbos & Ton, 2023). One key reason for this is that most people do not have a bicycle available on the activity-side of the train trip. Taking the bicycle onto the train can be cumbersome and expensive or sometimes simply not possible, while having a second bicycle parked on the activity-side can be expensive and also causes spatial problems in the highly congested bicycle parking garages. In recent years, shared micromobility (SMM) services have appeared as a possible solution. The Dutch Railways introduced their OV-fiets (PT-bike) service back in 2003 and in 2023, travellers made 5.9 million trips with bicycles available at over 300 stations nation-wide. Despite this success, the work of Jonkeren & Huang (2024) highlights that the potential to increase the number of activity-end trips performed by SMM remains high.

To summarise the wide range of studies on (shared) micromobility, Abduljabbar et al. (2021) and Zhu et al. (2022) both carried out reviews of literature on the topic, looking into the perception of SMM, willingness to use it and its environmental impact. Their key findings are that SMM can help alleviate congestion, improve accessibility and reduce emissions. They both mention the benefit of improving access/egress to PT, yet they both point to a lack of studies analysing the level of integration and the benefits in relation to it. Several studies conclude that the users of SMM are primarily younger, male, highly educated, with a higher income and living in highly urbanised areas (Aguilera-García et al., 2020; Badia & Jenelius, 2023; Christoforou et al., 2021; Mitra & Hess, 2021; Reck & Axhausen, 2021; Shelat et al., 2018). Chahine et al. (2024) clustered users based on two different topics, one based on how SMM benefits are perceived and another on the barriers of adopting SMM. The largest cluster (78%) on the topic of benefits were individuals who acknowledge the existence of SMM benefits, while not adopting it, perhaps because they do not see it as something for them. At the same time, the largest barrier cluster (61%) are indifferent about the barriers, not really considering them, likely because they have no intention of using SMM and are thus not informed. In terms of attribute importance, Chahine et al. (2024) cite safety, reliability, health and convenience as the most important among all groups.

Clustering is a common approach of segmenting the population to better understand the individual needs and perceptions of various subgroups. Alonso-González et al. (2020) and van 't Veer et al. (2023) both used a latent class clustering analysis (LCCA) to study the perception and potential adoption of Mobility-as-a-Service (MaaS), while van der Meer et al. (2023) used it to study the potential use and adoption of neighbourhood mobility hubs. Their conclusions show similar results with two larger clusters (25-35% each), where one seems to be ready to adopt new technologies, while the other is more conservative. Among the smaller segments, they find a highly enthusiastic group, already using a variety of travel modes, including many shared services and a highly averse, very negative group, somewhat older group. While not directly analysing SMM, the results of these studies are interesting as MaaS shares many similarities with SMM, also since SMM is a key component of MaaS.

Acceptance and adoption and of new services is often characterised by a variety of factors and not all may play an equally important role for everyone. Perhaps the most well-known and broadly used is the Theory of Planned Behaviour (Ajzen, 1991), outlining three broad factors affecting intention to use, namely the attitude, perception of others (subjective norm) and ease-of-use (perceived behavioural control). A more recent framework, extending on the factors we see in the Theory of Planned Behaviour is the Unified Theory of Acceptance and Use of Technology 2 (UTAUT2) model (Venkatesh et al., 2012), also including the expected performance, hedonic motivation, habit etc. Later studies adopting it have also added constructs, based on the technology being studied. Van 't Veer et al. (2023) and van der Meer et al. (2023) for example both used the UTAUT2 framework to construct attitudinal statements on the topics of MaaS and Mobility hubs respectively, adding additional constructs as they saw fit.

Although many studies have analysed different aspects of shared micromobility (SMM) and applied different clustering approaches, to the best of our knowledge, none has looked at the integration of SMM with public transport or the drivers and barriers associated with it. Our study fills this gap by using an established framework, namely the UTAUT2, to develop attitudinal statements and constructs to aid in the understanding of how different user groups perceive shared micromobility in combination with public transport and what are the key drivers and barriers of adoption of the individual groups. Through these insights, we provide policy recommendations on how to make SMM more attractive to different user groups. The rest of paper is structured as follows. The data collection and data analysis methods are outlines in Section 2, with the results of the data analysis presented in Section 3. A discussion on the observed attitudes and behaviour of survey respondents is thereafter discussed in Section 0, with the conclusions and recommendations presented in Section 5.

# 2 Methodology

In this section, we outline the methods used and data collected for the purposes of our study. To start off, we define and operationalise the conceptual framework in Section 2.1. Once the data for the conceptual framework is collected, we estimate an exploratory factor analysis to narrow down the number of indicators into constructs, as outlined in Section 2.2. This data is then used to cluster users based on their attitudes and perceptions by means of a latent class cluster analysis. This step is further expanded upon in Section 2.3. Finally, we describe the data collection process, including the sample size, statistics, representativeness and other questions asked of the respondents in the survey. This last step is outlined in Section 2.4.

## 2.1 Conceptual framework and attitudinal statements

To study the perception of SMM, we adapt the UTAUT2 framework and develop the associated attitudinal statements to measure the individual constructs, the list of which can be seen in Table 1. We take six of the original constructs from Venkatesh et al. (2012) and add an additional four constructs, namely *Reliability*, *Perceived risk*, *Sustainability* and *Health*, which were found to be highly important for the topic of shared mobility by Chahine et al. (2024) and van 't Veer et al. (2023). *Reliability* in the context of SMM refers primarily to vehicle availability for the traveller. As there tend to be a limited number of vehicles available, or especially if it is a free-floating system (vehicles can be rented and parked anywhere), users have no guarantee on having a vehicle waiting for them. *Perceived risk* mainly refers to physical safety, as some users report not feeling safe on specific SMM modes, for example due to their higher speed or having to use the road. *Sustainability* is a key selling point of SMM and research has shown that individuals who adjust their behaviour to be more sustainable (more environmentally conscious) tend to be more likely to use SMM (van 't Veer et al., 2023). *Health* is another motivating factor or barrier for users in relation to SMM. Some see it as a healthy was of travel, by having to exercise for example, while others may not see it as such. To capture the relation of these constructs with behaviour intent, we also pose statements regarding the respondents intention to use SMM. The number of items measuring each construct is listed in Table 1, with the full set of attitudinal statements presented in Appendix A.

*Table 1. List of framework constructs, the number of items specified for each and source*

| Construct | Items | Source |
|---|---|---|
| Performance expectancy | 3 | (Venkatesh et al., 2012) |
| Effort expectancy | 6 | (Venkatesh et al., 2012) |
| Social influence | 5 | (Venkatesh et al., 2012) |
| Facilitating conditions | 6 | (Venkatesh et al., 2012) |
| Hedonic motivation | 4 | (Venkatesh et al., 2012) |
| Habit | 5 | (Venkatesh et al., 2012) |
| Reliability | 3 | (Chahine et al., 2024) |
| Perceived risk | 3 | (Chahine et al., 2024; van 't Veer et al., 2023) |
| Sustainability | 3 | (van 't Veer et al., 2023) |
| Health | 6 | (Chahine et al., 2024; van 't Veer et al., 2023) |
| Behavioural intention | 5 | (Venkatesh et al., 2012) |

The adjusted framework applied in this study is based on the work of van 't Veer et al. (2023), where the ten constructs are split between extrinsic motivation and intrinsic motivation. Extrinsic motivation can be defined as all the constructs relating to external factors influencing the perception of SMM and that can be influenced by other parties, i.e. the operator, friends, family etc. Conversely, intrinsic motivation is rooted deeper in the beliefs and values of the individual, for example what do they enjoy doing (hedonic motivation), how habitual is their behaviour or what is their personal opinion on the health and sustainability in general or regarding SMM specifically.

On top of that, our framework includes moderators, i.e. personal characteristics which may influence how people perceive SMM. Specifically, we include a variety of socio-demographic (age, gender, income, education,…) and travel behaviour (mode use frequency, mode preferences for different trip purposes, experience with SMM,…) characteristics which add information on the individuals and help in explaining their attitude and behaviour. The full list of questions regarding the socio-demographics and travel behaviour is shown in Appendix B. A graphic representation of our final adjusted UTAUT2 framework is presented in Figure 1.

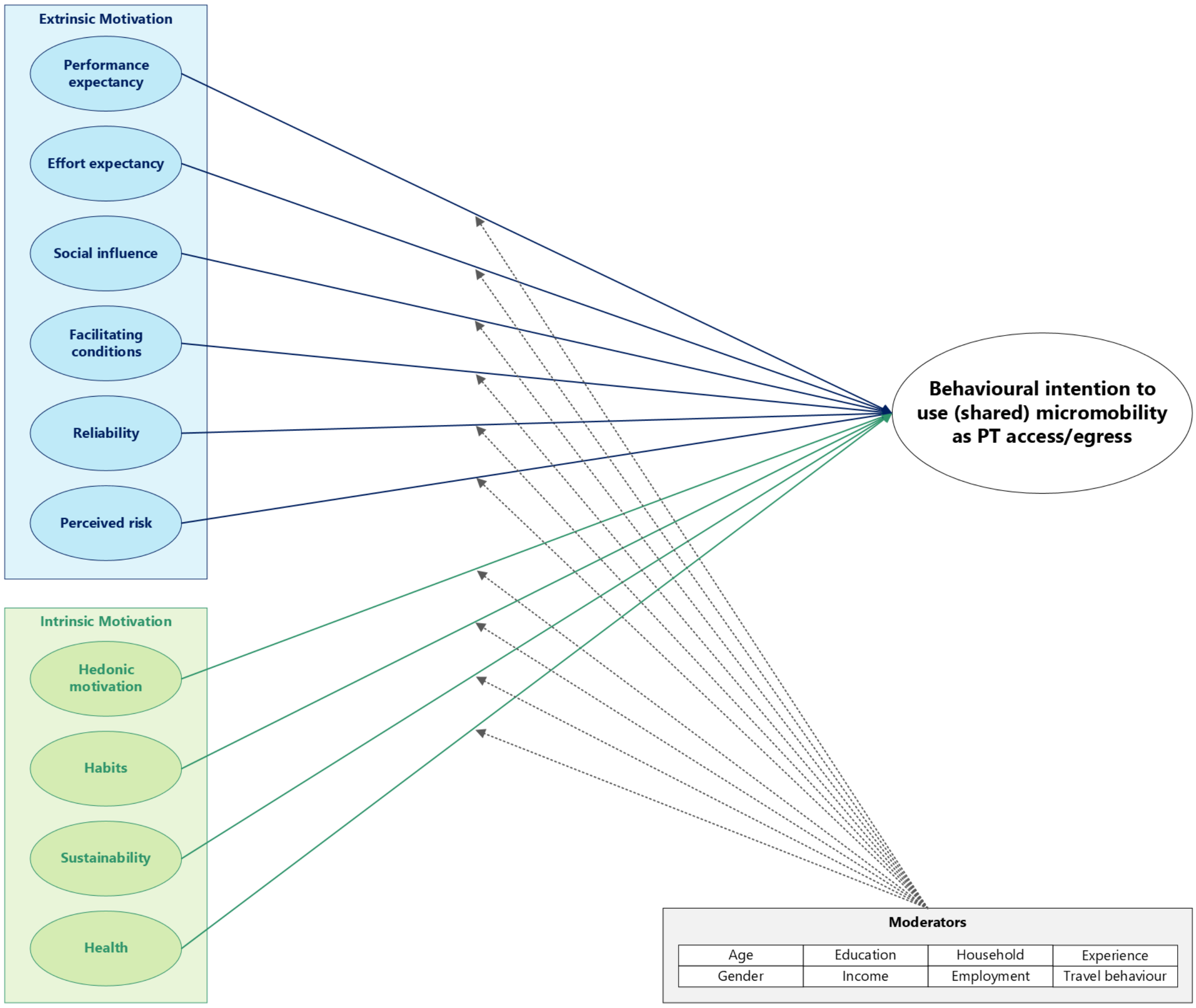

*Figure 1. Adjusted UTAUT2 framework*

## 2.2 Exploratory factor analysis

To obtain factors from the individual statements, we carry out an exploratory factor analysis (EFA). The first step of an EFA is checking if the data is suitable for the analysis. A common statistic for assessing this is the Kaiser-Meyer-Olkin (KMO) measure. The KMO value can be between 0 and 1, with a higher value indicating the data is more appropriate for EFA (Schreiber, 2021). We also compute Bartlett's test of sphericity and the determinant of the correlation matrix. These tests make sure that the data is correlated enough to extract factors, but not too correlated to cause multicollinearity issues. Hence, Bartlett' tests needs to be significant ($p < 0.05$), while the matrix determinant needs to be higher than 0.00001 (Field, 2013). If these criteria are not met, remedial action needs to be taken.

To extract the factors, we apply the maximum likelihood method. The factor loadings are extracted so that they are most likely to reproduce the correlation matrix. The number of extracted factors is based on the Kaiser rule, i.e. the factors which have an eigen value above 1 (Schreiber, 2021). Factors are rotated using an oblique method, specifically the oblimin technique. Oblique techniques allow for factor correlations, whereas orthogonal rotations do not (Schreiber, 2021).

Finally, we check factor loadings, cross-loadings and communality. Ideally, individual items should load highly onto one factor and low on all others (cross-load). Field (2013) considers factors loads above 0.3 acceptable, while Stevens (2001) states this should be based on the sample size, with samples over 1,000 respondents only requiring a loading of 0.162. For cross-loading, Taherdoost (2016) advises that values above 0.4 are unacceptable, while Samuels (2017) states it should be no higher than 75% of the main factor loading. For communality, Child (2006) suggests to only keep items with values above 0.2. In our analysis, we consider the factor loads to be acceptable if above 0.3 (Field, 2013), cross-loads acceptable based on both criteria, meaning below 0.4 (Taherdoost, 2016) and at the same time a most 75% of the main loading (Samuels, 2017), and communality above 0.2 (Child, 2006).

Once the EFA calculations are finalised, we interpret the meaning of each identified factor, based on which items load onto it. In some instances, for clarity and to avoid using negative or double negative phrases, we invert the factors by swapping the sign of the final factor score (from + to – or vice-versa).

## 2.3 Latent class cluster analysis

Once the EFA is concluded, the factor values for each respondent are calculated. This is then the start for the LCCA. We start by determining the ideal number of classes. We do this using only indicators (factors). The covariates (moderators) are added later, once the ideal number of classes is determined (van der Meer et al., 2023). To determine the number of classes, we assess the BIC value and the bivariate residuals (BVR). BIC is a measure combining model fit and the number of parameters, measuring the efficiency of the model, achieving a high model fit with as few parameters as possible. The best model is the one with the lowest BIC value (Vermunt & Magidson, 2005). BVR is a measure of remaining covariation between two factors. A value above 3.84 indicates the covariation is statistically significant and thus an additional cluster may be able to capitalise on this covariation (Schreiber, 2017). As many models show decreasing BIC even with a large number of classes, we can also consider the percentage improvement of BIC value as the cutoff point (Alonso-González et al., 2020; van der Meer et al., 2023).

Once the optimal number of classes is determined, we add all the covariates and then conduct a backwards elimination. We iteratively remove covariates that are insignificant ($p < 0.05$) until only significant ones are left. Insignificant covariates are kept as inactive to aid in cluster interpretation.

## 2.4 Data collection

The survey, with questions as outlined in Section 2.1 with the posed questions and attitudinal statements outlined in Appendices A and B, was operationalised through the Qualtrics survey tool. Data was collected through two different panels, namely the Dutch Railways own panel (NS Panel) (NS, 2020) and a commercial panel maintained by PanelClix. The NS Panel is used for its wide reach among existing train users. PanelClix on the other hand is included to also reach occasional and infrequent train users and to obtain a representative (sub)sample of the Dutch population. Data from both was collected in summer of 2024, with the NS panel data collected between the 30th of July and 31st of August, leveraging a total of 2,393 responses, while the PanelClix data was collected between the 26th and 30th of August, yielding 611 responses.

The data is the filtered, removing responses that did not consent to their data being stored and incomplete responses. Next, we check for straightlining behaviour. This is where respondents reply with the same answer to all attitudinal statements, even when this is completely illogical, as some questions are reverse coded. Finally, we remove responses that are deemed too fast to be realistic. This includes all responses that were faster than two standard deviations from the median response duration (Qualtrics, 2024). This leaves us with 1,371 responses from the NS panel and 520 from PanelClix, or a total of 1,891 valid responses to our survey.

An overview of the sample(s) characteristics and the population is presented in Table 2. We can see that overall, the PanelClix subsample is quite well representative of the population as a whole. There is a slight underrepresentation of older individuals (65+), those with a lower (elementary) education. Accordingly, middle-aged individuals (especially 35-49), those with a middle (vocational) education are overrepresented. Individuals with a driver's license are also somewhat overrepresented in the sample, whereas no clear conclusions can be made for income, due to the fairly high share of those not wishing to disclose their income.

On the other hand, according to Table 2, the NS panel sample is not very representative of the Dutch population. The NS panel was not used with a representative population in mind, but rather to get insights into the behaviour of existing train travellers. And although no definitive data exists on the Dutch train travelling population, the NS panel is often used as a proxy. As we see in Table 2, the NS panel sample tends to be older, with a higher income and very highly educated, more aligned with the Dutch train travelling population. Car ownership and consequently driving license ownership are also lower.

*Table 2. Socio-demographic characteristics of the two samples and the population*

| | | NS Panel | PanelClix | Population* |
|---|---|---|---|---|
| Gender | Man | 52% | 48% | 50% |
| | Woman | 48% | 52% | 50% |
| Age | 18-34 | 10% | 26% | 27% |
| | 35-49 | 23% | 28% | 22% |
| | 50-64 | 33% | 27% | 25% |
| | 65+ | 33% | 19% | 25% |
| Household size | One person | 30% | 19% | 19% |
| | Multiple people | 70% | 81% | 81% |
| Work status | Working | 63% | 69% | 67% |
| | Not working | 37% | 31% | 33% |
| Education level | Low | 4% | 17% | 29% |
| | Middle | 20% | 53% | 36% |
| | High | 75% | 30% | 35% |
| Income | Low | 8% | 18% | 20% |
| | Middle | 44% | 48% | 45% |
| | High | 30% | 21% | 35% |
| | n/a | 18% | 13% | - |
| Driving license | No | 16% | 9% | 20% |
| | Yes | 84% | 91% | 80% |
| Car ownership | Average | 0.79 | 1.29 | 1.11 |

** the population characteristics are based on the >18 population*

With this dual sample, we are able to assess both the preferences of existing users and of the potential new users. All models are estimated on the full sample to leverage the large number of responses we obtained. However, the cluster presentations are accompanied by both the sample and population characteristics. What we from here on refer to as population refers to the PanelClix subsample which, as we have shown is quite well representative for the Dutch 18+ population.

# 3 Results

In this section, we present the process of applying the EFA and LCCA methodologies and their outcomes, as described in Sections 2.2 and 2.3 respectively. The obtained factors are presented in Section 3.1 with a detailed overview of the population segments discussed in Section 3.2.

## 3.1 Exploratory factor analysis

We use SPSS software to perform the EFA and start with all 48 statements. The full dataset can be considered meritorious, with a KMO value of 0.89 (Schreiber, 2021). The dataset also fulfils Bartlett's test, with a $p < 0.01$. However, the determinant of the correlation matrix is too low ($1.3 \cdot 10^{-9}$), one item has a communality below 0.2 (social_2) and four items have unacceptably high cross-loadings. To remedy these issues, we remove items that do not have a sufficiently high loading, have too high cross-loads or a too low communality and re-estimate an EFA. This process is repeated iteratively (removing individual items that do not satisfy the criteria, until all the aforementioned statistics are satisfied and the determinant achieves an acceptable level ($\geq 10^{-5}$).

Through several iterations, we achieve an acceptable model, retaining 25 of the 48 items, loading onto eight factors (down from 12 in the initial full model). The new KMO value is 0.84, meaning the data is still meritorious, Bartlett's test is still significant, the matrix determinant is acceptable ($1.13 \cdot 10^{-5}$). All communalities are above 0.3, all items have a loading of at least |0.4| and only one cross-loading of 0.246 (facility_1), which passes both criteria of cross-loading, that they should be below 0.4 and at most 75% of the main loading (0.550 in the case of facility_1). The eight factors explain 73% of the variability. The final model can be seen in Table 3.

*Table 3. Final EFA model, with 25 items loading onto eight factors*

| Items | Factors | | | | | | | |
|---|---|---|---|---|---|---|---|---|
| | 1 | 2 | 3 | 4 | 5 | 6 | 7 | 8 |
| intention_1 | 0.937 | | | | | | | |
| intention_2 | 0.800 | | | | | | | |
| intention_3 | 0.850 | | | | | | | |
| intention_4 | 0.479 | | | | | | | |
| reliability_1 | | -0.699 | | | | | | |
| reliability_2 | | -0.977 | | | | | | |
| sustainability_1 | | | -0.882 | | | | | |
| sustainability_2 | | | -0.773 | | | | | |
| sustainability_3 | | | -0.668 | | | | | |
| social_1 | 0.655 | | | | | | | |
| social_3 | | | | 0.924 | | | | |
| social_4 | | | | 0.851 | | | | |
| effort_1 | | | | | 0.619 | | | |
| effort_3 | | | | | 0.862 | | | |
| effort_4 | | | | | 0.664 | | | |
| health_4 | | | | | | -0.850 | | |
| health_5 | | | | | | -0.768 | | |
| hedonic_2 | | | | | | | -0.832 | |
| hedonic_4 | | | | | | | -0.857 | |
| risk_1 | | | | | | | -0.635 | |
| facility_1 | | | | | *0.246* | | | 0.550 |
| facility_2 | | | | | | | | 0.605 |
| facility_3 | | | | | | | | 0.655 |
| facility_5 | | | | | | | | 0.722 |
| habit_5 | | | | | | | | 0.589 |

Next, we interpret the eight factors to better understand what they are portraying. Some factors have negative signs (F2, F3, F6 and F7), meaning that items load negatively onto them. Additionally, items loading onto F4 are phrased in a negative way ("bad social image"). Given this, the five factors are inverted, easing the interpretation. The names of the eight factors are listed below, with the factors being inverted shown in *italic*:

1. Intend to use SMM
2. *Confident about SMM vehicle availability*
3. *Climate conscious*
4. *SMM has a good societal image*
5. SMM is easy to use
6. *Using PT is a healthy way of travel*
7. *Mopeds are a fun and safe way of travel*
8. Confident with using (digital) technology

## 3.2 Latent class cluster analysis

In the following step, we perform the LCCA, by estimating models with up to ten classes. As shown in Table 4, the best fitting model given BIC is the 9-class model. Given the % change in BIC, a 4 or 5-class model seem to fit better as the model fit improvements are minor afterwards. However, looking at the BVR, we notice a big change with the 6-class model, after which BVR does not change drastically. BVR assesses the level of covariation between factors and values over 3.84 indicate there is still covariation which can be capitalised on with additional classes. Using this combination of metrics, and also evaluating the interpretability of classes, we decide to continue with the 6-class model.

*Table 4. Overview of the number of classes and associated model fits*

| # Classes | BIC | % change BIC | max(BVR) | min(class size) |
|---|---|---|---|---|
| 1 | 40,834 | -5.68% | 660 | 100% |
| 2 | 38,513 | -2.60% | 387 | 40% |
| 3 | 37,511 | -1.37% | 168 | 23% |
| 4 | 36,999 | -0.95% | 145 | 15% |
| 5 | 36,646 | -0.54% | 104 | 10% |
| 6 | 36,449 | -0.81% | 41 | 8% |
| 7 | 36,153 | -0.37% | 45 | 9% |
| 8 | 36,021 | -0.83% | 40 | 8% |
| 9 | 35,721 | 0.20% | 31 | 8% |
| 10 | 35,792 | -5.68% | 36 | 4% |

We add the socio-demographic and travel behaviour information as covariates and iteratively remove insignificant parameters, changing them into inactive covariates. Only three socio-demographic characteristics remain among the active covariates, namely the age, gender and income. The majority of active covariates are travel related: number of cars in the household, train subscriptions, frequency of bicycle use, experience using shared bicycles, experience with other shared mobility services, preferred mode for commuting, the frequency of using train for work trips and frequency of using train for shopping trips.

Based on the model outcomes, the cluster characteristics and their attitudes towards SMM, we give each class a name. We also provide the size of each cluster based on both the sample and the population. This latter step is done by utilising the PanelClix subsample which is deemed representative of the population, as outlined in Section 2.4. This is done by using the cluster allocation probabilities of the PanelClix respondents and calculating the number falling into each of the six clusters. All the names and cluster sizes are presented in Table 5. Afterwards, the attitudes and characteristics of all six clusters are presented. To aid us in this, the factor scores (deviations from the population average) are presented, also in Table 5, with a graphic representation (also including the sample deviation) in Figure 2.

Additionally, the average factor scores per cluster with respect to the sample average are shown in Table 9 in Appendix C. Next, the preferred access mode to train stations shown in Figure 3, respondents weekly travel patterns and use of different modes on a weekly basis in Figure 4 and Figure 5 respectively, as well as their experience using shared mobility services shown in Figure 6. Finally, the socio-demographic characteristics are showcased in Table 6.

Cluster 1, the biggest in the sample, is called **Shared mobility positives (C1)**. They show the second highest intention to use SMM, are digitally savvy and climate conscious. They have the strongest belief that SMM is viewed positively in society and, interestingly, they are the only ones in the sample who view it more positively than negatively (F4). On SMM vehicle availability, fun and safety, they are average for the sample. As the biggest cluster in the sample, they do not stand out strongly on many socio-demographic characteristics, however they tend to be younger, highly educated and with a high income. Compared to the population, they tend to have an above average income, education level, more likely to have train travel subscriptions and less likely to own a private car. They also tend to be some of the most experienced shared bicycle users. In terms of travel behaviour, they use most modes, although are less likely to use the private car. Accessing the train station on their home-end, they are much more likely to use the bicycle compared to any other mode.

The second cluster we term the **Car-oriented sharing neutrals (C2)**. As the second biggest (20%) in the sample and the biggest in the population (35%), they show mainly opposing, but mild views to the *Shared mobility positives*. They are less climate conscious and think SMM has a bad social connotation. They also do not think SMM is easy to use. Interestingly, they are more likely to see it as fun compared to other clusters. They tend to be the most representative with respect to the population in almost all socio-demographic and travel-related characteristics, yet compared to the sample they tend to have a lower level of education (no university degree) and have a low-to-middle income. They have the highest average household car ownership, translating into the highest car use of any cluster, with over 2/3 using it on a weekly basis and are also the most likely to use car to access the train station. They tend to be less experienced with using shared bicycles, while being equal to most other clusters when it comes to other shared modes.

Next is a cluster we label as **Older apprehensive sharers (C3)**. Like the *Shared mobility positives*, they are concerned about the climate. They are somewhat positive on their intention to use SMM, possibly because they see the added value of it, while also thinking it is difficult to use, and not as fun and safe. They are also not as confident using smartphones and have a neutral opinion about the public perception of SMM. On average, they are the oldest of the clusters, and thus also the most pensioner-dominated cluster. They are highly educated and also the most female-dominated cluster. They are much more likely to live in a household without kids. Looking at travel behaviour, they have the lowest car ownership and the highest likelihood of having a train travel subscription. They are the least likely of any cluster to travel by car to the train station. They have above average experience with the shared bicycle (OV-fiets), but below average experience with other shared modes.

The most enthusiastic cluster is the fourth, namely the **Young eager adopters (C4)**. They show some of the strongest attitudes of any cluster. With the highest intention to use SMM, highest confidence in SMM vehicle availability, and strongest climate awareness. They are confident in using SMM, find it exciting, fun and safe, and are highly tech savvy. Interestingly, they do think SMM makes them look bad among their friends and family, but they likely do not care or do not find it important. They are the most male-dominated at over 2/3, the youngest and with a high income. Like the *Car-oriented sharing neutrals*, they are likely to live with children and have more than one car in the household. They are one of the clusters that travels most, with all available modes. Surprisingly, they also stand out among holders of other train travel subscriptions, including peak-time discounts, meaning they also travel a lot by train. Their tech-savviness also translates into being the most experienced cluster when it comes to using shared mobility and other shared services.

The fifth cluster are the **(Shared) Mobility avoiders (C5)**. In terms of their attitudes, they are the most negative and thus the most opposite to the *Young eager adopters*. They show the lowest intention to use SMM, find it dangerous, difficult to use and are concerned about its availability. They are also the most climate indifferent, although they do not necessarily see SMM having a negative societal connotation. Like the *Older apprehensive sharers*, this cluster tends to be older and more female. They are also the lowest educated and with a lower income. They have the highest share that are neither working nor retired. They have an above average share of stay-at-home partners and those unable to work. They have a fairly low car ownership and are some of the most likely to travel with public transport, specifically local public transport (bus, tram, metro), for example when travelling from their home to the train station. They are the least experienced with shared mobility or shared services out of any cluster. They are also more likely to not travel much at all.

Finally, we turn to the sixth cluster, the **Skilled sharing sceptics (C6)**. They do not show strong positive or negative tendencies towards adoption of SMM, are confident they would not have difficulty using SMM and are digitally savvy. They tend to be middle-aged and with a high income and an average education profile. They are the most likely to be working, with 71% employed. They also share their above average car ownership and high car use with the *Young eager adopters*, but travel comparatively less with public transport and more by car. They are however fairly well versed in using a variety of shared services, often coming in second or third, just after the *Young eager adopters* and sometimes after the *Shared mobility positives*.

*Table 5. Clustering model outcomes, with average factor deviation from the population for each cluster.*
*(Red/Dark green indicate a strong negative/positive relationship while Orange/Light green indicate a mild negative/positive relationship)*

| | | Shared mobility positives | Car-oriented sharing neutrals | Older apprehensive sharers | Young eager adopters | (Shared) Mobility avoiders | Skilled sharing sceptics |
|---|---|---|---|---|---|---|---|
| | Cluster size in population | 22% | 35% | 8% | 14% | 12% | 9% |
| | Cluster size in sample | 35% | 20% | 17% | 10% | 10% | 8% |
| | Factors | | | | | | |
| F1 | Intent to use SMM | 0.35 | 0.00 | 0.16 | 1.14 | -1.42 | 0.16 |
| F2 | Confident about SMM vehicle availability | -0.42 | -0.04 | -0.44 | 0.66 | -0.94 | -0.22 |
| F3 | Climate conscious | 0.80 | -0.27 | 0.85 | 0.87 | -0.40 | 0.24 |
| F4 | SMM has a good societal image | 1.24 | -0.34 | 0.20 | -0.20 | 0.32 | 0.16 |
| F5 | SMM is easy to use | 0.51 | -0.20 | -0.45 | 0.68 | -0.88 | 0.45 |
| F6 | Using PT is a healthy way of travel | 0.16 | 0.02 | 0.40 | 0.68 | -0.12 | -0.02 |
| F7 | Mopeds are a fun and safe way of travel | -0.50 | 0.02 | -0.96 | 0.55 | -1.40 | 0.01 |
| F8 | Confident with using (digital) technology | 0.28 | -0.16 | -0.45 | 0.82 | -1.15 | 0.40 |

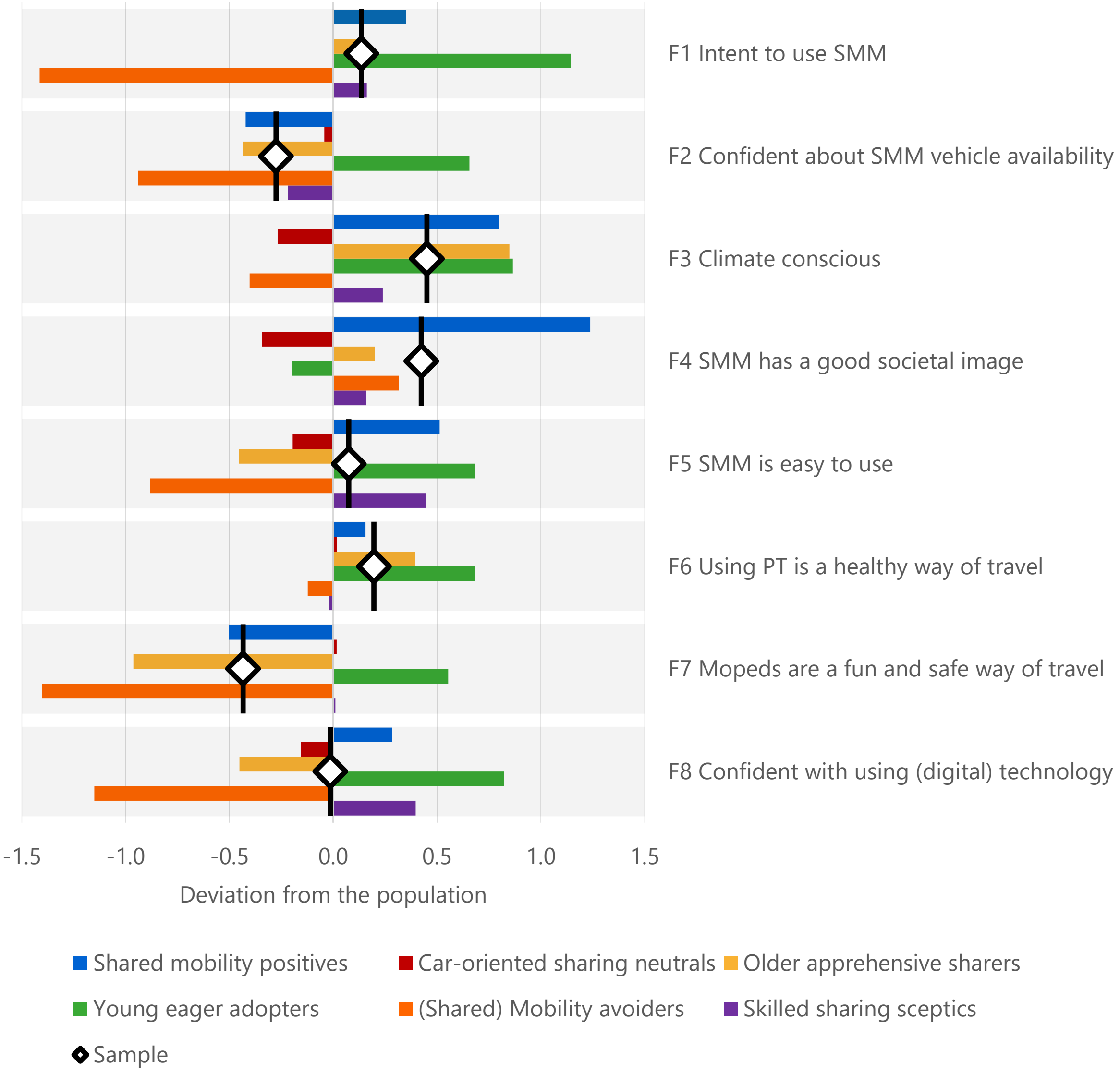

*Figure 2. Deviations of the cluster averages and the sample from the population*

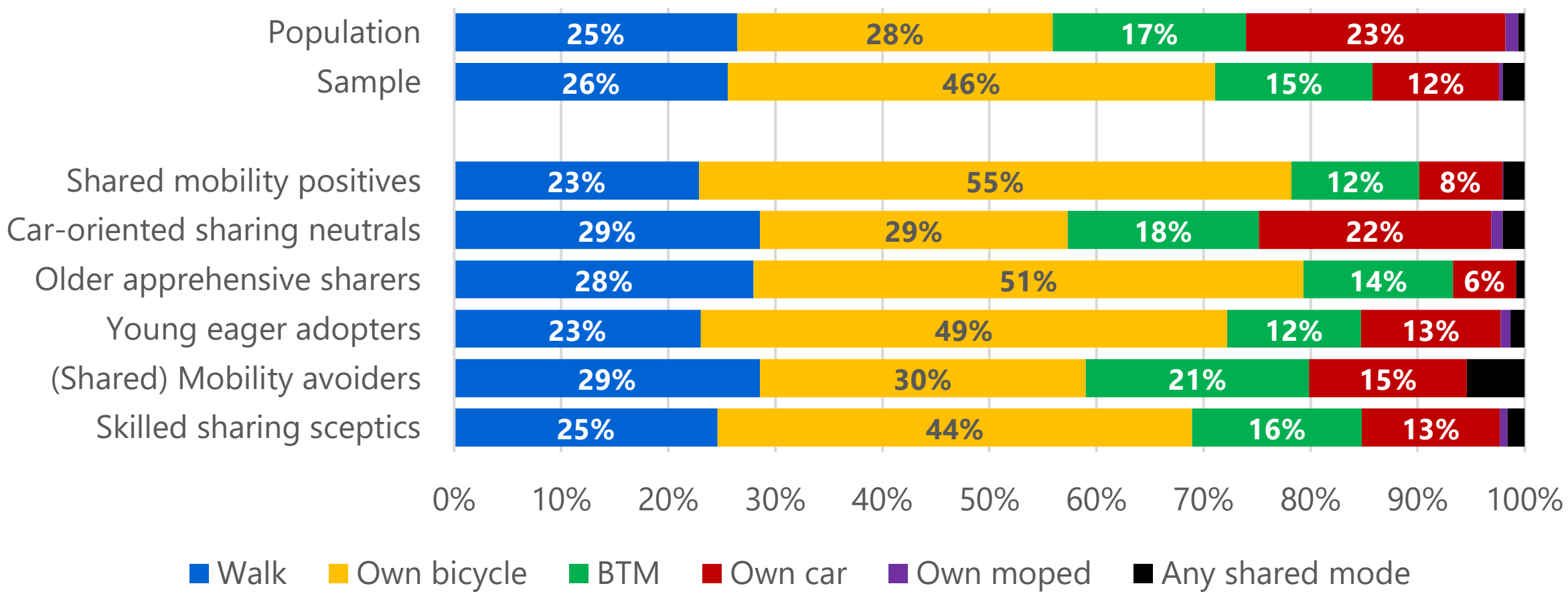

*Figure 3. Preferred mode for accessing the train station on the home-end*

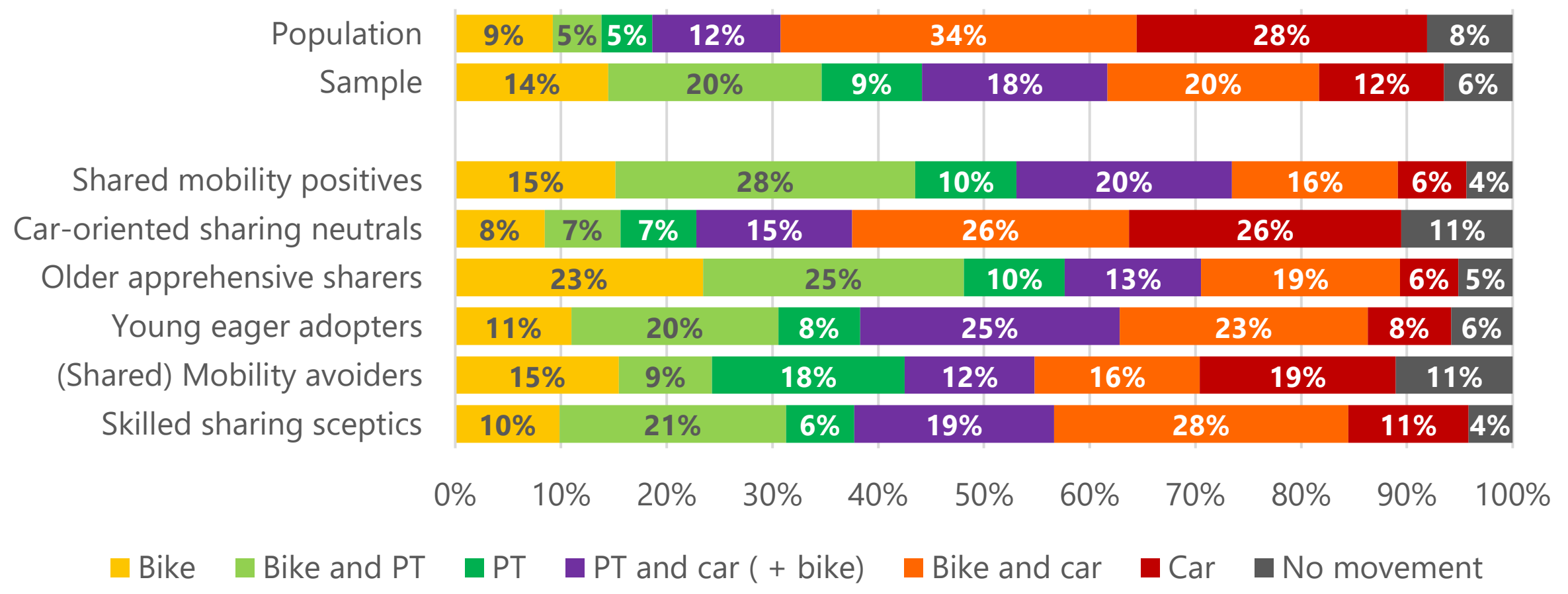

*Figure 4. Modes (or mode combinations) used on a weekly basis for each cluster*
*(in mode combinations, it means that all modes are used at least once on a weekly basis, not necessarily at part of the same trip)*

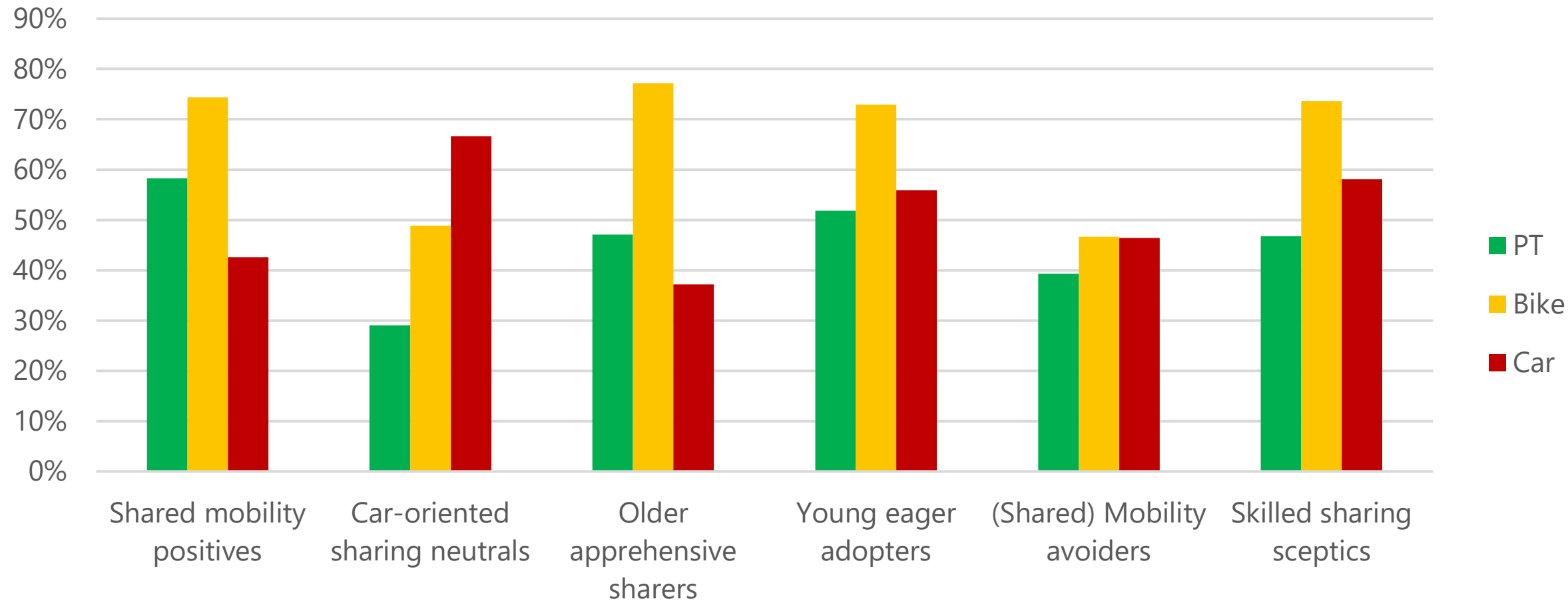

*Figure 5. What share of cluster members use different modes on a weekly base*

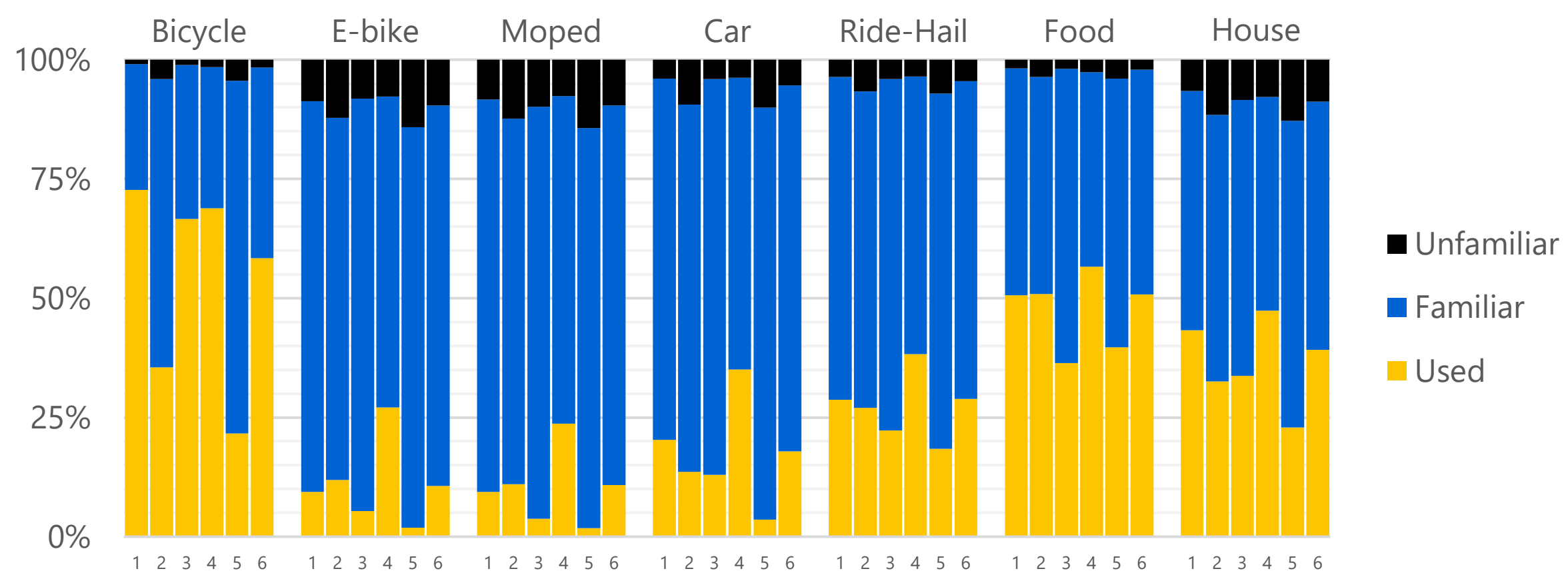

*Figure 6. Experience with various shared services*

*Table 6. Socio-demographic characteristics of each cluster. Green text indicates values above the sample mean, while red text indicates values below the sample mean*

| | | Population | Shared mobility positives | Car-oriented sharing neutrals | Older apprehensive sharers | Young eager adopters | (Shared) Mobility avoiders | Skilled sharing sceptics |
|---|---|---|---|---|---|---|---|---|
| Cluster size in population | | | *22%* | *35%* | *8%* | *14%* | *12%* | *9%* |
| Cluster size in sample | | | *35%* | *20%* | *17%* | *10%* | *10%* | *8%* |
| Gender | Female | 48% | 52% | 45% | **61%** | **32%** | 55% | 41% |
| | Male | 52% | 48% | 55% | **39%** | **68%** | 45% | 59% |
| Age | 18-34 | 26% | **13%** | 18% | **5%** | 21% | **8%** | **14%** |
| | 35-49 | 28% | 27% | 25% | **15%** | 24% | 22% | 25% |
| | 50-64 | 27% | 35% | 26% | 28% | 31% | 29% | 33% |
| | 65+ | 19% | 26% | **30%** | **53%** | 24% | **42%** | 27% |
| Education | Low | 17% | **4%** | 18% | **5%** | 9% | 16% | **6%** |
| | Middle | 53% | **22%** | 43% | **21%** | **31%** | **37%** | **31%** |
| | High | 30% | **74%** | 38% | **74%** | **60%** | **46%** | **62%** |
| Income | Low | 18% | 8% | 15% | 10% | 10% | 16% | **6%** |
| | Middle | 48% | 45% | 47% | 45% | 49% | 40% | 41% |
| | High | 21% | **35%** | 18% | 22% | **34%** | 17% | **39%** |
| | n/a | 13% | 13% | 20% | 22% | 8% | **27%** | 13% |
| Work status | Working | 69% | 63% | 65% | **48%** | 67% | **47%** | 71% |
| | Retired | 15% | 15% | 18% | **38%** | 13% | **27%** | 14% |
| | Other | 16% | 22% | 18% | 14% | 20% | 26% | 15% |
| Household | Single | 19% | 22% | 23% | **31%** | 19% | **33%** | 22% |
| | Couple (no kids) | 38% | 39% | 36% | 46% | 36% | 43% | 41% |
| | With kids | 33% | 23% | 27% | 15% | 27% | 13% | 25% |
| | Other | 10% | 15% | 13% | 8% | 18% | 11% | 12% |
| Cars in household | 0 | 12% | **33%** | 15% | **36%** | **23%** | **34%** | **24%** |
| | 1 | 55% | 53% | 55% | 55% | 50% | 53% | 50% |
| | 2+ | 34% | **13%** | 30% | **9%** | 26% | **13%** | 26% |
| | *mean* | 1.29 | 0.83 | 1.23 | 0.73 | 1.11 | 0.80 | 1.05 |
| Train subscription | None | 70% | **46%** | 65% | **28%** | **39%** | **51%** | 67% |
| | Off-peak | 12% | **39%** | 20% | **59%** | **31%** | **35%** | 22% |
| | Other | 18% | 15% | 15% | 14% | **30%** | 14% | 12% |

***Green*** *indicates 10% points or more* ***above*** *the sample average*
***Red*** *indicates 10% pointes or more* ***below*** *the sample average*

# 4 Discussion on attitudes and behaviour

Beyond the attitudinal clustering exercise outlined in previous chapters, comparing stated attitudes with stated choice behaviour provides critical insight into the consistency between what people believe and how they act. The survey collected both attitudinal statements and responses to a stated choice experiment on the topic of SMM integration with rail-based PT, with the latter analysed in our previous work (Geržinič et al., 2025). Our segmentation analysis based on stated behaviour creates a valuable methodological contrast to our attitudinal clustering in the current study. Analysing the relationship between these two approaches and identifying potential overlap between clusters is essential for developing a comprehensive understanding of the attitude-behaviour gap and strengthening the validity of behavioural prediction models in this domain.

Both the LCCA approach used in this study, as well as the latent class choice model (LCCM) methodology applied by Geržinič et al. (2025) cluster individuals probabilistically to one of the emergent segments. Based on these allocation probabilities, we construct a contingency table with the number of respondents belonging to each of the class-cluster combinations. We use the term class when referring to the LCCM method performed by Geržinič et al. (2025) and the term cluster to refer to the segmentation carried out through the LCCA in this paper. The contingency table with the shares of respondents in each class-cluster combination is shown in Table 7, with the full table with the number of respondents in each presented in Table 10 in Appendix D. To check for potential correlation between class and cluster membership, we perform a chi-square test, the details of which are outlined in Appendix D. It shows that there are indeed correlations between classes and clusters, since the null hypothesis is rejected.

Looking closer at the outcomes in Table 7, we see quite some expected outcomes based on the analyses of both studies, but also some unexpected correlations. There seems to be a positive correlation between the *Sharing hesitant cyclists* and *Older apprehensive sharers*, both showing behaviour and attitudes of wishing to use SMM and having some experience with it, but also worries about other aspects. Another strong and expected positive correlation is between *(Shared) Mobility avoiders* and *Sharing-averse PT users*. Both segments are highly negative about SMM and sharing in general, showing limited intention of using it due to the complexity and perceived unsafety. Those who tend to have a *Young eager adopters* mindset are more likely to behave as *Multimodal sharing enthusiasts*, which is another expected correlation, since both are open to sharing, have plenty of experience with it and see themselves as capable.

On the other hand, it is somewhat surprising to see a correlation between *Car-oriented sharing neutrals* and *Multimodal sharing enthusiasts*, given the fairly strong aversion to sharing by the former and openness to it by the latter. What may be linking them is their above average car ownership and car use when compared to other segments in their respective analyses. Given the large size of the *Multimodal sharing enthusiasts* class (by far the largest of three) in the study by Geržinič et al. (2025), it is also not surprising that multiple clusters from this study fall within it. Additionally, two interesting groupings are *Shared mobility positives – Sharing hesitant cyclists* and the negative correlation between the *Skilled sharing sceptics* and *Sharing-averse PT users*. The former correlation is interesting, as it was expected that the *Shared mobility positives* would correlate highly with the *Multimodal sharing enthusiasts*, given the openness towards sharing and multimodal mindset of both. The negative correlation of the latter is also interesting, as they both see limited added value in SMM and sharing in general, however their drastically different capabilities in using shared services and digital technologies is what may be the cause of this negative correlation.

*Table 7. Comparison of class and cluster membership from the current work and work by* Geržinič et al. (2025)

| | Sample | Multimodal sharing enthusiasts | Sharing hesitant cyclists | Sharing-averse PT users |
|---|---|---|---|---|
| Sample | | **43%** | **24%** | **33%** |
| Shared mobility positives | *34%* | 42% | **27%** | 30% |
| Car-oriented sharing neutrals | *19%* | **49%** | **17%** | 34% |
| Older apprehensive sharers | *17%* | **33%** | **30%** | **37%** |
| Young eager adopters | *11%* | **49%** | 24% | **28%** |
| (Shared) Mobility avoiders | *10%* | **38%** | **17%** | **45%** |
| Skilled sharing sceptics | *8%* | 46% | 24% | **30%** |

***Green*** *indicates membership that is more than 10%* ***above*** *expected*
***Red*** *indicates membership that is more than 10%* ***below*** *expected*

What is interesting from this analysis is that although the chi-square test was strongly significant, the alignment between attitudinal segments and behavioural clusters is less pronounced than might be anticipated. Specifically, we found that attitudinal segments and behavioural clusters only partially overlapped, rather than showing a near-complete correspondence that would be expected if attitudes directly translated to behaviours. This points to the fact that there is often a disconnect between people's opinions/attitudes and their actual behaviour. This phenomenon can be referred to as cognitive dissonance, with several studies reporting on it in the transportation domain. De Vos & Singleton (2020) carried out a literature review on the topic of cognitive dissonance in transportation and found that people's attitudes towards certain modes often do not match their behaviour. Similar findings were reported by An et al. (2022), having studied attitudes and behaviour among the Dutch population. Also when focusing on shared mobility, users tend to exhibit paradoxical behaviour with respect to attitudes and behaviour (Magnani & Re, 2020) with Magnani et al. (2018) also showing that while users show enthusiasm for shared mobility, they are still reluctant to use it.

While most research suggests that the behavioural aspect is more reliable when attitudes and behaviours do not align, it is important to recognize that in our context, both analyses rely on hypothetical scenarios rather than revealed preferences, making the comparison less definitive. Furthermore, our attitudinal study and the behavioural study outlined in our other study (Geržinič et al., 2025), examine complementary aspects of SMM-PT integration: our current research explores broader opinions, beliefs, and attitudes, while our previous work investigates specific econometric trade-offs that travellers consider. Therefore, the observed differences between clustering approaches are not so much concerning as they are helpful. The meaningful correlations that do exist between the two segmentation exercises demonstrate their interconnectedness, while the variations highlight that they cannot be treated as interchangeable. This underscores the importance of studying both attitudinal and behavioural aspects to develop a deeper understanding of the problem at hand.

# 5 Conclusion

In this research, we study how different user groups perceive shared micromobility (SMM) in combination with public transport and what are the key drivers and barriers of adoption of the individual groups. By performing an exploratory factor analysis (EFA), we obtain eight factors relating to different aspects of SMM such as safety, ease-of-use, societal perception and pleasure. Additionally, we get information on respondent's attitudes towards public transport, climate change and digital savviness. Further, we carry out a latent class cluster analysis (LCCA), resulting in six clusters, namely the *Shared mobility positives*, *Car-oriented sharing neutrals*, *Older apprehensive sharers*, *Young eager adopters*, *(Shared) Mobility avoiders* and *Skilled sharing sceptics*. The most polarising factors are on SMM ease-of-use, fun and safety related to E-moped use and the social image of SMM. The least polarising is if travel by PT is considered healthy or not.

Analysing at the individual clusters, it is interesting to consider what would motivate each of them to use/try SMM, allowing us to better understand their needs and what operators/policymakers could do to encourage the use of SMM. Starting with the most excited, the *Young eager adopters*, they do not seem to need any additional encouragement, as they are the most likely to already be using SMM services. The next are the *Shared mobility positives*: their main barrier to wider adoption is vehicle availability and the danger/stress of using e-mopeds. While not scoring particularly strongly on either factor, these are the most worrying aspects in their eyes, likely making them the strongest barriers to broader use. The social image, ease of use and technological savviness do not seem to be perceived as barriers by this group. Next are the *Skilled sharing sceptics*, who's main issue with SMM is likely the bad connotation associated with its use and the vehicle availability. While the latter can be addressed through different operational strategies, the former requires broader societal discussions on the topic and sometimes also sufficient time for people to accept such novelties. *Older apprehensive sharers* (scoring very similarly on the intention to use SMM), find SMM fairly difficult to use and also a dangerous and stressful experience. The former may likely be due to their lower tech savviness, meaning that help from personnel and having non-digital options to rent vehicles would be beneficial. Their awareness of climate issues may also help stimulate them to try SMM. A similar issue with high technological dependence can be observed among the *Car-oriented sharing neutrals*. In addition to this, they also associate SMM with very negative perception in their social circles, meaning wider acceptance of such services would be needed for them to consider it. Finally, the *(Shared) Mobility avoiders* would likely be the last group to adopt SMM, finding almost all aspects as a barrier, from social perception, ease of use, danger, vehicle availability, etc.

Our analysis uncovers vehicle availability as the primary barrier to wider train-SMM adoption. This concern affects even groups likely to use the service, including *Shared mobility positives*, *Skilled sharing sceptics*, and *Older apprehensive sharers*. For groups with a lower adoption likelihood, like *Car-oriented sharing neutrals* and *(Shared) Mobility avoiders*, ease-of-use and digital skill requirements present additional obstacles. Non-digital options and staff assistance would also benefit *Older apprehensive sharers* who want to use SMM but lack the confidence. Societal perception remains challenging and requires policymakers to clearly communicate benefits while improving regulations, ensuring orderly operations, promoting equitable access, and developing proper infrastructure to transform SMM's public image from nuisance to valuable mobility solution.

While providing valuable insights, the study and its outcomes also have certain limitations. As any stated preference approach, the stated adoption likelihood may have been overstated by the respondents. This is somewhat mitigated by including questions on respondents' revealed behaviour, although further studies verifying the adoption of SMM should be carried out. Additionally, SMM is made up of many different modes, not all of which could be captured here. We therefore recommend additional studies investigating other SMM modes. Linking the findings of the study at hand and our previous work (Geržinič et al., 2025), we uncover potential cognitive dissonance among individuals, where their stated attitudes and behaviour seem contradicting. While this specific topic was not a point of interest of this study, it does open interesting future avenues of research to gain a better understanding of how and why people develop attitudes and how they do or do not act on them. Finally, while care was taken to include all socio-demographic groups and a closely representative subsample was collected, we cannot be certain of its full representativeness.

## Acknowledgements

This study is part of a larger project investigating the perceptions and adoption potential of shared micromobility as a train station access/egress mode. Other studies within this project investigate the stated and revealed choice behaviour, uncovering the respondents' willingness-to-pay and the associated market segmentation and traveller heterogeneity.

We would like to thank the Dutch Railways (NS) for their help and cooperation in this research, but by co-funding the research project as well as providing access to their invaluable in-house panel of train travellers, which was used to gather data for this study.

# Appendices

## A. Attitudinal statements

Attitudinal statements are developed based on the constructs defined in the UTAUT2 framework (Venkatesh et al., 2012), which is a frequently used and cited technology use and acceptance model. We adjust the constructs and develop 3-6 statements for each of the constructs. The full list of constructs is provided below:

1. Performance expectancy
   1.1. I believe that using shared micromobility will save me time when travelling.
   1.2. I believe that using shared micromobility will make my travel less efficient than it is now.
   1.3. I believe that using shared micromobility will save me money.
2. Effort expectancy
   2.1. I expect it will be easy for me to learn how to use a shared (electric) bicycle.
   2.2. I expect it will be easy for me to learn how to use a shared electric moped.
   2.3. I believe I will not have problems unlocking shared (electric) bicycles on my own.
   2.4. I believe I will not have problems unlocking shared electric mopeds on my own.
   2.5. I think it is easier to use shared micromobility if the vehicles are all parked together in the same location.
   2.6. I think it is difficult to find information on how to use shared micromobility (sign-up, create an account, unlock a the vehicle,…).
3. Social influence
   3.1. I can see myself using shared micromobility.
   3.2. My public image (how people see me) is important to me.
   3.3. My friends would think less of me if I used shared micromobility.
   3.4. My family would think less of me if I used shared micromobility.
   3.5. I believe it is societally responsible to use shared micromobility.
4. Facilitating conditions
   4.1. I have a smartphone. (move to socio-demographics)
   4.2. I know how to use smartphone applications.
   4.3. I have smartphone applications for (one or more) travel companies on my smartphone.
   4.4. I do not mind having multiple different applications for different travel companies on my smartphone.
   4.5. I would prefer unlocking shared micromobility vehicles using a card (e.g. OV chipkaart) and not a smartphone application.
   4.6. I do not mind making payments through smartphone applications.
5. Hedonic motivation
   5.1. It is fun to use a shared (electric) bicycle.
   5.2. It is fun to use a shared electric moped.
   5.3. I can enjoy my surroundings when I travel by (electric) bicycle.
   5.4. I can enjoy my surroundings when I travel by electric moped.
6. Habit
   6.1. I would need to make big changes to my travel pattern to start using shared (electric) bicycles or electric mopeds.
   6.2. I tend to use the same mode of transport when travelling.
   6.3. I tend to use the same route when travelling.
   6.4. I am open to trying new products and services.
   6.5. I am open to trying new digital applications.

7. Reliability
    7.1. I am confident that there will always be a shared vehicle available at the station.
    7.2. I am confident that there will always be a shared vehicle available for my return trip to the station.
    7.3. I am willing to pay more to have the certainty of having the shared vehicle for the entire round trip (leaving the station and coming back after the activity).
8. Perceived risk
    8.1. I feel safe when riding an electric moped.
    8.2. I feel safe when travelling by public transport in the Netherlands.
    8.3. I feel safe when riding an (electric) bicycle.
9. Sustainability
    9.1. I am concerned about the effects of climate change.
    9.2. I am aware of the impact transport has on climate change.
    9.3. I have adjusted my travel behaviour due to the impact it has on the climate.
10. Health
    10.1. I believe walking is a healthy way of travelling
    10.2. I believe cycling is a healthy way of travelling
    10.3. I believe that using electric vehicles (electric bicycle or moped) is a healthy way of travelling.
    10.4. I believe that using bus/tram/metro is a healthy way of travelling.
    10.5. I believe that using the train is a healthy way of travelling.
    10.6. I take health benefits of different modes into account when making travel choices.
11. Behavioural intention
    11.1. I intend to use shared micromobility services when travelling by train
    11.2. I intend to use shared micromobility when going to work or education.
    11.3. I intend to use shared micromobility when visiting friends/family.
    11.4. I would travel by train more if I had more shared mobility options to get to/from the station.
    11.5. I would travel by train more if I had more public transit (e.g. bus/tram/metro) options to get to/from the station.

## B. Socio-demographic & Travel behaviour questions in the survey

In addition to attitudinal statements, pertaining directly to the UTAUT2 constructs, we also collect data to assess the moderators of the model. For socio-demographic data, we collect the following characteristics:

- Gender
- Age
- Highest level of completed education
- Annual household income
- Employment status
- Household composition
- Number of cars in the household
- Smartphone ownership
- Driving license ownership

For data relating to travel behaviour, we collect the information as outlined in Table 8. The table also includes the available answers, to clearly indicate how the data collection was operationalised.

*Table 8. Travel behaviour questions and possible answers*

| Questions | Answers |
|---|---|
| How frequently do you make use of your | • 4 or more days per week |
| • **Own car**<br>• **Own (E-)bicycle**<br>• **Bus/Tram/Metro**<br>• **Train** | • 1-3 days per week<br>• 1-3 days per month<br>• 6-11 days per year<br>• 1-5 days per year<br>• (almost) never |
| How many trips did you make with the airplane last year? *1 trip consists of an outbound and return flight and any potential transfers* | • 6 or more trips<br>• 3-5 trips<br>• 1-2 trips<br>• none |
| What is the mode you use most for the following trips? *If you use multiple modes, please select the one with which you travel most (longest distance).*<br>• **For work or education**<br>• **Visiting friends and family**<br>• **Recreation / Sports**<br>• **Shopping** | • Car<br>• Bicycle<br>• Walk<br>• Bus/Tram/Metro<br>• Train<br>• Other (moped, scooter, rollerblades,…) |
| How often do you travel by train for the following trips?<br>• **For work or education**<br>• **Visiting friends and family**<br>• **Recreation / Sports**<br>• **Shopping** | • 4 or more days per week<br>• 1-3 days per week<br>• 1-3 days per month<br>• 6-11 days per year<br>• 1-5 days per year<br>• (almost) never |
| When travelling by train, how do you most often get to the station from your home? | • Walk<br>• Own Car<br>• Own (E-)Bicycle<br>• Shared car/bike/moped<br>• Bus/Tram/Metro<br>• Other |
| When travelling by train, how do you most often get from the station to your destination?<br>• **For work or education**<br>• **Visiting friends and family**<br>• **Recreation / Sports**<br>• **Shopping** | • Walk<br>• Own car<br>• Own (E-)Bicycle<br>• Shared car<br>• Shared bike<br>• Shared moped<br>• Bus/Tram/Metro<br>• Other |
| How much experience do you have with the following shared modes and sharing economy services?<br>• **Shared car**<br>• **Sared bicycle (OV fiets)**<br>• **Shared e-bike (OV e-bike)**<br>• **Shared moped (Check, Felyx, Go)**<br>• **Ride-hailing (Uber, Via, Lyft)**<br>• **House sharing (airbnb)**<br>• **Food delivery (Thuisbezorgd, UberEATS)** | • Never heard of it / Didn't know it existed<br>• I know it but never tried/used<br>• Do not want to try it<br>• Would like to try it<br>• Used it once<br>• Used a few times<br>• Use it regularly |

## C. Factor scores in the sample

Average factor scores for each of the six clusters. Unlike the results in Table 5 in Section 3.2, where the scores show the deviation from the estimated population average (based on a representative subsample of the population), the factor scores in Table 9 show the direct outcomes of the factor scores, based on the whole sample on which the model was estimated.

*Table 9. Clustering model outcomes, with average factor values for each cluster. Red/Dark green indicate a strong negative/positive relationship while Orange/Light green indicate a mild negative/positive relationship*

| Factors | | Clusters | | | | | |
|---|---|---|---|---|---|---|---|
| | | C1 | C2 | C3 | C4 | C5 | C6 |
| F1 | Intent to use SMM | 0.21 | -0.14 | 0.02 | 1.01 | -1.55 | 0.03 |
| F2 | Confident about SMM vehicle availability | -0.15 | 0.23 | -0.16 | 0.93 | -0.66 | 0.06 |
| F3 | Climate conscious | 0.35 | -0.72 | 0.40 | 0.41 | -0.85 | -0.21 |
| F4 | SMM has a good societal image | 0.81 | -0.77 | -0.22 | -0.62 | -0.11 | -0.26 |
| F5 | SMM is easy to use | 0.44 | -0.27 | -0.53 | 0.61 | -0.96 | 0.37 |
| F6 | Using PT is a healthy way of travel | -0.04 | -0.18 | 0.20 | 0.49 | -0.32 | -0.22 |
| F7 | Mopeds are a fun and safe way of travel | -0.07 | 0.45 | -0.53 | 0.99 | -0.97 | 0.44 |
| F8 | Confident with using (digital) technology | 0.30 | -0.14 | -0.44 | 0.84 | -1.14 | 0.41 |

## D. Chi-square test

In order to perform the chi-square test, we construct a contingency table of the respondents and observe the frequency of belonging to different class-cluster combinations, as shown in Table 10. Next, we calculate the expected frequencies. This is done by taking the sizes of the classes and clusters as a whole and simply multiplying them to obtain the expected size of each combination. This is shown in Table 11.

The null hypothesis in the chi-square test states that there is no correlation between the two tables, meaning that the differences between observed and expected frequencies should be insignificant in order to confirm the null hypothesis. Performing the test with 10 degrees of freedom, we see that in fact the p-value is $2.43 \cdot 10^{-12}$, meaning the difference is highly significant and that there in fact are correlations between the two analyses and the null hypothesis cannot be accepted.

*Table 10. Observed frequency of respondents belonging to the different class-cluster combinations*

| | Multimodal sharing enthusiasts | Sharing hesitant cyclists | Sharing-averse PT users |
|---|---|---|---|
| Shared mobility positives | 271 | 177 | 196 |
| Car-oriented sharing neutrals | 181 | 62 | 125 |
| Older apprehensive sharers | 108 | 97 | 121 |
| Young eager adopters | 99 | 49 | 56 |
| (Shared) Mobility avoiders | 72 | 32 | 86 |
| Skilled sharing sceptics | 74 | 38 | 47 |

*Table 11. Expected frequency of respondents belonging to the different class-cluster combinations*

| | Multimodal sharing enthusiasts | Sharing hesitant cyclists | Sharing-averse PT users |
|---|---|---|---|
| Shared mobility positives | 274 | 155 | 215 |
| Car-oriented sharing neutrals | 157 | 89 | 123 |
| Older apprehensive sharers | 139 | 79 | 109 |
| Young eager adopters | 87 | 49 | 68 |
| (Shared) Mobility avoiders | 81 | 46 | 64 |
| Skilled sharing sceptics | 67 | 38 | 53 |